\documentclass{INTERSPEECH2023}
\usepackage{caption}
\usepackage{subcaption}
\usepackage{xcolor}

\interspeechcameraready

\title{Improved DeepFake Detection Using Whisper Features}
\name{Piotr Kawa$^1$, Marcin Plata$^1$, Micha{\l} Czuba$^1$, Piotr Szymański$^1$, Piotr Syga$^1$}
\address{
  $^1$Wroc{\l}aw University of Science and Technology, Wroc{\l}aw, Poland
}
\email{\{piotr.kawa, marcin.plata, michal.czuba, piotr.szymanski, piotr.syga\}@pwr.edu.pl}

\begin{document}

\maketitle

\begin{abstract}
With a recent influx of voice generation methods, the threat introduced by audio DeepFake (DF) is ever-increasing. Several different detection methods have been presented as a countermeasure.
Many methods are based on so-called front-ends, which, by transforming the raw audio, emphasize features crucial for assessing the genuineness of the audio sample.
Our contribution contains investigating the influence of the state-of-the-art Whisper automatic speech recognition model as a DF detection front-end.
We compare various combinations of Whisper and well-established front-ends by training 3 detection models (LCNN, SpecRNet, and MesoNet) on a widely used ASVspoof 2021 DF dataset and later evaluating them on the DF In-The-Wild dataset. 
We show that using Whisper-based features improves the detection for each model and outperforms recent results on the In-The-Wild dataset by reducing Equal Error Rate by 21\%.
\end{abstract}

\section{Introduction}

Audio DeepFakes (DF) is a collection of deep learning techniques that create artificial speech. These methods may involve creating entirely new sentences using Text-To-Speech or Voice-Cloning (aiming at mimicking speech patterns of a specific person or sounding \textit{natural} to a human listener) or transferring the qualities of the victim's voice to the attacker's speech, referred to as Voice-Conversion~\cite{survey-df}. With the increasing sophistication of deep learning techniques, it has become relatively easy to create audio DeepFakes that are difficult to distinguish from bona fide recordings. Such malicious activities can cause significant harm, including compromising the security of systems protected by speaker recognition and contributing to spreading fake news or defaming an individual's reputation. As a result, developing effective DF detection techniques has become increasingly critical for ensuring the integrity and trustworthiness of audio-based communication systems. This need has been reflected in the recent growth of methods based on deep neural networks that assess the validity of utterances~\cite{lcnn,specrnet,mesonet,rawnet2,aasist,wav2vec_aasist}.

Detecting DeepFake audio is a problem analogous to speech spoofing~\cite{spoofing1,spoofing2}. Despite the superficial similarity, they differ in the targets they deceive: spoofing aims to fool speaker verification systems, whereas DF targets humans. One can easily point out attacks typical for one of the areas (e.g., replay attack for spoofing) that are not present in the other --- therefore, these areas are considered separate.

Audio feature extraction is crucial in many applications, like speech recognition or speaker identification. Many existing approaches to DeepFake detection focus on some extracted features instead of a raw waveform. That makes the extraction method vital to DF detection and generates motivation for in-detail investigation.  Feature extraction aims to identify an audio signal's key characteristics and emphasize them. Mel-frequency cepstral coefficients (MFCC)~\cite{mfcc} and linear-frequency cepstral coefficients (LFCC)~\cite{lfcc} are some of the most widely used methods. MFCCs are based on the human auditory system's non-linear frequency response. In contrast, LFCCs are designed to address some of the limitations of MFCCs, such as their insensitivity to low-frequency information and lack of robustness to noise. 

Whisper~\cite{Whisper} is a state-of-the-art automatic speech recognition (ASR) system. It was trained on 680,000 hours of content. Due to the data's diversity and magnitude, it has shown to be robust against broad spectra of background interferences, accents, and languages. Its name refers to the family of the models differing, i.a., by width and sizes of layers.
Whisper is based on off-the-shelf encoder-decoder Transformer architecture~\cite{attention_is_all_you_need}. 
Its encoder is based on two convolutional layers, each processed by a GeLU activation function~\cite{gelu}. The information is later modified by adding position embeddings~\cite{attention_is_all_you_need}. The encoder ends with a series of the pre-activation residual attention blocks~\cite{generating_long_sequences}, followed by the normalization layer.

In this work, we harness the feature extraction capabilities of the pre-trained Whisper's encoder not to capture speech properties later used for ASR but to investigate its performance in DF detection.
We use it along with three detection models to verify if neural network-provided features might help in DF detection. 
We selected Whisper for the evaluation due to its effectiveness in speech recognition, which comes from the large and diversified speech corpora that it was trained on. As such, we infer that Whisper's features would ignore most of the naturally occurring artefacts and help identify artificially modified speech samples. In particular, it would help with the problem of generalization, which refers to the poor efficacy of the models on the data outside of the training set's distribution --- currently one of the most challenging problems in DF detection.
Our experiments cover the smallest available Whisper version --- \textit{tiny.en}. We aim at minimal overhead to ensure that the provided solution can be widely used in production environments. In addition, the model was trained strictly on English data, which is the main language among DF detection datasets.
Please note that bigger Whisper versions were shown to yield even more satisfying results in various tasks (e.g. \textit{large} performs up to 3x better~\cite{Whisper} in speech recognition or translation than the 
\textit{tiny.en} model). This allows us to expect that the results can be enhanced even further.
Note that Whisper was trained on human speech samples --- a bona fide and thus highly-biased set in the sense of DF detection. However, a similar approach was proposed in~\cite{wav2vec_aasist}. The authors used a front-end based on wav2vec 2.0~\cite{wav2vec_2} that was originally designed as an unsupervised pre-trained model for representations and used in the task of speech recognition. wav2vec was also trained on data considered bona fide. Fine-tuning this front-end led to a substantial increase in the results and generalization~\cite{wav2vec_darts,vicomtech}.
We choose Whisper for our approach as it significantly improves over wav2vec, not only in the results reported but also in the data scale used for training and self-supervision (over 16x more). While this may play a lesser role for speech recognition as the ratio of importance between the audio model and language model can differ between approaches, we treat Whisper as an audio encoder only and thus expect to see an impact of a much larger dataset used in training.

The codebase related to our research can be found on GitHub: github.com/piotrkawa/deepfake-whisper-features.

\section{Detection models and datasets}\label{sec:models}

We consider four models --- three processing spectrogram-like features: LCNN~\cite{lcnn}, MesoNet~\cite{mesonet} (MesoInception-4 variant), and SpecRNet~\cite{specrnet} as well as RawNet3~\cite{rawnet3} that analyzes raw audio.
The models consist of respectively 467,425 (LCNN), 28,486 (MesoNet), 277,963 (SpecRNet), and 15,496,197 (RawNet3) parameters. 
SpecRNet used in our comparison differs from its original implementation --- to enable processing of the higher-dimensions front-ends we add an adaptive 2D average pooling after the last SeLU layer~\cite{selu}.
A similar scenario occurred in MesoNet, where we add adaptive 1D average pooling right before the penultimate fully connected layer. In the case of LCNN, we increase the size from 160 to 768 of input features and hidden features of two bi-LSTM layers and the input features of the last Linear layer.

For the spectrogram-based models, we consider 3 front-ends: LFCC, MFCC, and the output of the Whisper ASR encoder. 
We additionally evaluate the concatenated front-ends of cepstral-coefficients with Whisper features. The intuition behind it follows~\cite{aad} --- concatenation of different front-ends may yield better results.
We use LFCC and MFCC based on the window and hop lengths of 400 and 160; they are composed of 128 coefficients. We concatenate front-ends in a second dimension with its delta and double delta features. This results in the shape data (128 * 3, 3000).
In our experiments, we use \textit{tiny.en} variant of the Whisper model. Its encoder contains 7,632,384 parameters and outputs data of shape (376, 1500). To match it with the size of the other front-ends (required for using it in the concatenated front-end setting), we replicate one of the dimensions achieving a tensor of size (376, 3000).

The dataset used in the paper consists of 125,000 samples randomly selected from ASVspoof 2021 (DF)~\cite{asvspoof_2021} and all 31,779 samples of DeepFakes In-The-Wild~\cite{wild}. The decision is motivated by the general scarcity of DF datasets, of which ASVspoof is among the largest and most popular. In contrast, the latter dataset consists of samples reflecting real-world scenarios (being gathered from the Internet). To emulate the scenario in which architectures are developed using training on the most popular datasets, they should be effective in the actual environment while determining the authenticity of new samples, possibly distorted by noise.
Even though several models achieve high efficacy on popular datasets like ASVspoof or WaveFake~\cite{wavefake}, the investigation in~\cite{wild} showed that those methods do not generalize well to unknown, real-world samples. The EER of LCNN evaluated on the In-The-Wild dataset increased by up to 1000\%, and even more for RawNet3. Naturally, one of the solutions might be similar to the one presented in~\cite{aad}, mixing the datasets used so that multiple creation methods may be recognized. Such an approach may be infeasible in practical scenarios, where new DF creation methods should also be detected. To mimic the practical scenario in our investigation, we decided to train the models on a well-established DF dataset, as an end-user would, and later test it on the dataset that reflects a possible real-world sample to verify.

\section{Experimental setup}\label{sec:benchmark_description}

Each sample underwent a standardized preprocessing procedure. It covered resampling to 16~kHz mono-channel, removing silences that were longer than 0.2~s and padding (by repetition), or trimming samples to 30~s of content. For two reasons, we decided on the input length of the 30~s instead of the typical length of about 4~s (~\cite{rawnet2,aad,wavefake, specrnet}). Firstly, works like~\cite{wild} show that analyzing longer utterances yields better results. Secondly, Whisper takes as an input 30~s of content: samples are trimmed or padded with zeros. Instead of padding it with zeros, we decided to fill the whole input tensor with speech.

We trained models on a random subset of 100,000 training and 25,000 validation samples of the ASVspoof 2021 DF dataset. We used a subset of this dataset for two reasons: we wanted to make our solution possible to train on a single GPU in about 24 hours; moreover --- we did not anticipate a significant gain from the samples' quantity for an architecture like Whisper \textit{tiny}. We addressed the disproportion between bona fide and fake classes with oversampling. 
We used a learning rate of $10^{-4}$ and a weight decay of $10^{-4}$ for all spectrogram-based models. RawNet3 used a learning rate of $10^{-3}$ and a weight decay of $5\cdot 10^{-5}$. We trained models with a binary cross-entropy function for 10 epochs with a batch size of 8. Training of RawNet3 included SGDR scheduling~\cite{sgdr} with a restart after each epoch.
The checkpoint of the highest validation accuracy was selected for later tests on the full In-The-Wild dataset. We present our results using Equal Error Rate (EER) metric as a fraction. EER is commonly used in DF and spoofing problems. To ensure reproducibility, we ran each process with a fixed randomness seed. Each experiment was run on a single NVIDIA TITAN RTX GPU (24GB VRAM).

\section{Benchmarks}\label{sec:benchmark}

\subsection{Baseline comparison}\label{sec:baseline_benchmark}

Our baseline comparison covered LCNN, MesoNet, SpecRNet and RawNet3 models. We tested front-ends of LFCC, MFCC and using Whisper's encoder. The encoder was not optimized (the weights were frozen), i.e., we used it purely as a feature extractor and based solely on its pre-trained features.

\begin{table}[h]    
    \centering
    \caption{The comparison of EER scores on In-The-Wild dataset. Using Whisper's encoder as a front-end contributes to the significant enhancement in case of SpecRNet and LCNN networks.}
    \begin{tabular}{c c c}
    \hline
        Model & Front-end & EER \\
        \hline
        SpecRNet & LFCC & 0.5184 \\
        SpecRNet & MFCC & 0.6897 \\
        SpecRNet & Whisper & 0.3644 \\
 
        LCNN & LFCC & 0.7756 \\
        LCNN & MFCC & 0.6762 \\
        LCNN & Whisper & 0.3567 \\

        MesoNet	& LFCC & 0.5451 \\
        MesoNet	& MFCC & 0.3132 \\
        MesoNet	& Whisper & 0.3856 \\

        RawNet3	& - & 0.5199 \\
        \hline
     \end{tabular}
    \label{tab:basic}
\end{table}

Note that the results presented in Tab.~\ref{tab:basic}, similarly to those reported in~\cite{wild}, deviate from the low errors typically reported in the DeepFake detection literature. This phenomenon is caused by training the models on an 'artificial' dataset created in a controlled manner (ASVspoof), whereas the evaluation is done on real-world samples from In-The-Wild dataset~\cite{wild}. The distribution of the artifacts in both sets differs significantly. As shown, the models do not have sufficient generalization capabilities, and when verified on recordings of a substantially different nature, the detection capabilities deteriorate significantly. Additionally note that, both, in the case of~\cite{wild} that was trained on LA subset of ASVspoof 2019~\cite{asvspoof_2019}, and in this paper (trained on DF subset of ASVspoof 2021), some of the models perform worse than random guessing (EER=0.5).
These results do not undermine the models in the traditional setup. In fact, LCNN with LFCC front-end and SpecRNet with MFCC, i.e., two architectures that achieved the worst results on In-The-Wild dataset, scored a satisfying EERs of 0.0149 and 0.0218 during validation on ASVspoof 2021. 
Notably, in the case of all detection models, LFCC and MFCC front-ends tend to provide features that were well-suited in the case of ASVspoof, and when trained and verified on the data from the same source have high efficacy~\cite{asvspoof_2021_2}, yet do not occur regularly in the DeepFakes from the In-The-Wild dataset. We discuss the nature of the extracted features in more depth in Sect.~\ref{sec:features_comparison}. 
Interestingly, smaller architectures -- SpecRNet and MesoNet, seem to generalize better and provide higher efficacy than LCNN. In fact, MesoNet (MFCC-variant) achieved the lowest EER. These results are similar to the ones reported in~\cite{wild}, where the model achieved the best results among the spectrogram-based architectures. One of the reasons may be the lower number of parameters, which results in a lesser degree of 'adjusting' towards the artifacts specific to the ASVspoof datasets, thus, higher generalization capabilities. 
Using Whisper-based features helps with generalizing. In the case of SpecRNet, we achieve a 29.71\% improvement in comparison with LFCC and 47.17\% in comparison with MFCC. In the case of LCNN, we improve both by 54\% and 47.25\%. Following the intuition that additional information may improve the detection, we investigate the synergy of the feature extractors in~Sect.~\ref{sec:joint_frontends}. Moreover, to improve the results even further and to address the worse results in the case of MesoNet, we decided to unfreeze the model (Sect.~\ref{sect:finetuning}).

\paragraph*{Constant Q-cepstral coefficients}

One of the popular spectrogram-based front-end used in speech and audio signal processing is Constant Q-cepstral coefficients (CQCC)~\cite{cqcc}. Works like~\cite{spoofing1,spoofing2} tested these features for spoofing detectors trained on ASVspoof 2021 (LA). To provide an extensive investigation of different front-ends, we used the CQCC with the LCNN model. However, when training the architecture with the same parameters as other feature extractors, the results on ASVspoof 2021 DF were unsatisfactory, achieving only around 60\% accuracy on train and test datasets. Consequently, we did not proceed with training other models using CQCC or evaluating them on the In-The-Wild dataset.

\subsection{Concatenated front-ends}\label{sec:joint_frontends}

Works like~\cite{aad} showed that using the concatenation of multiple front-ends could increase the detectors' effectiveness. The discussed pipeline did not differ from the one in Sect.~\ref{sec:benchmark}. We considered spectrogram-based models and used them with a concatenation of the \textit{classical} front-ends and Whisper's encoder. 

\begin{table}[h]
    \centering
    \caption{The EER values of the models using concatenated front-end features. The only enhancement is visible in case of two models, whereas for other models it had a negative impact.}
    \begin{tabular}{c c c}
    \hline
        Model & Front-end & EER \\
        \hline
        SpecRNet & Whisper + LFCC & 0.3485 \\
        SpecRNet & Whisper + MFCC & 0.4116 \\
        LCNN & Whisper + LFCC & 0.6270 \\
        LCNN & Whisper + MFCC & 0.6117 \\
        MesoNet & Whisper + LFCC & 0.8029 \\
        MesoNet & Whisper + MFCC & 0.3822 \\
        \hline
     \end{tabular}

    \label{tab:multi}        
\end{table}

\begin{table}[h]
    \centering
    \caption{Evaluation of the fine-tuned Whisper feature extraction (as a sole front-end and in concatenation, similarly as in Sect.~\ref{sec:joint_frontends}), on DeepFake In-The-Wild dataset. Note that 'EER (frozen)' refers to the case, where Whisper's encoder was not fine-tuned (cf. Tab.~\ref{tab:basic}, Tab.~\ref{tab:multi}) and 'EER (tuned)' to the results of fine-tuned extractor.}
    \begin{tabular}{c c c c}
    \hline
        Model & Front-end & EER (frozen) & EER (tuned) \\
        \hline
        SpecRNet & Whisper + LFCC & 0.3485  &  0.3795 \\
        SpecRNet & Whisper + MFCC & 0.4116 & 0.3769 \\
        SpecRNet & Whisper & 0.3644 & 0.3338 \\
        LCNN & Whisper + LFCC & 0.6270& 0.6270 \\
        LCNN & Whisper + MFCC & 0.6117 & 0.5899 \\
        LCNN & Whisper & 0.3567& 0.3290 \\
        MesoNet & Whisper + LFCC & 0.8029 & 0.5526\\
        MesoNet & Whisper + MFCC & 0.3822 & 0.2672\\
        MesoNet & Whisper & 0.3856 & 0.3362\\
            \hline
     \end{tabular}
    \label{tab:fine-tuned}
\end{table}

The comparison between the results of concatenated features (Tab.~\ref{tab:multi}) and a single feature extractor (cf.~Tab.~\ref{tab:basic}) shows that LCNN and SpecRNet models based on cepstral front-ends improve when trained with Whisper features. Detection enhances by up to 40.32\% for SpecRNet and up to 19.15\% for LCNN.
This suggests some positive synergy between the features and that additional knowledge is gained. However, the synergy does not guarantee that the results of joint-features detection outperform the Whisper-based. This may be caused by 'covering' some of the important Whisper features by the spectrogram-based front-ends. We investigate this issue in Sect.~\ref{sec:features_comparison}.
Conversely, one may notice a negative synergy between the Whisper-extracted features and the others (not a substantial one in the case of MFCC and significant in the case of LFCC). One may assume that the antagonistic effect is due to the architecture details and 'compression' of the provided information, which results in performing the classification based on the 'noised' information rather than an enhanced set of features. In Sect.~\ref{sec:features_comparison}, we show how the extracted features may be presented and discuss the issue further.

\begin{figure*}[t]
     \begin{subfigure}[b]{0.24\textwidth}
         \centering
         \includegraphics[width=\textwidth]{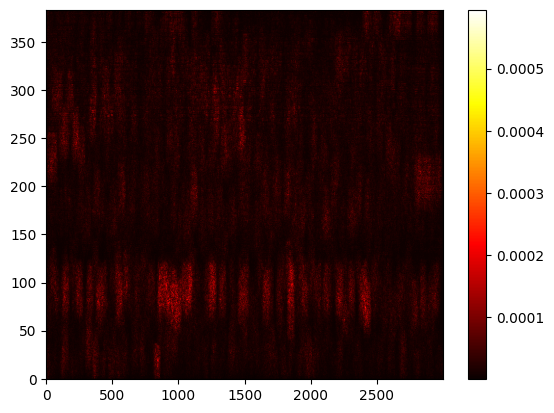}
         \caption{LFCC}
         \label{fig:meso_lfcc}
     \end{subfigure}
          \begin{subfigure}[b]{0.24\textwidth}
         \centering
         \includegraphics[width=\textwidth]{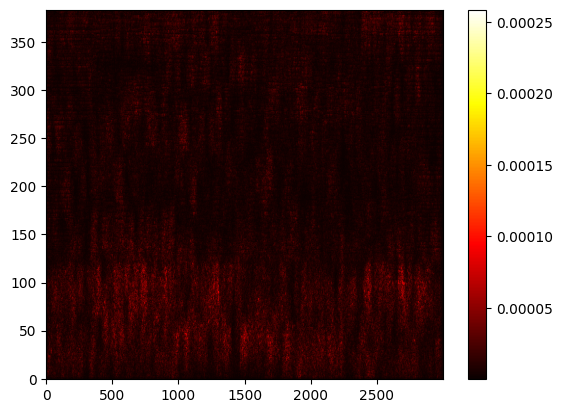}
         \caption{MFCC}
         \label{fig:meso_mfcc}
     \end{subfigure}
     \begin{subfigure}[b]{0.24\textwidth}
         \centering
         \includegraphics[width=\textwidth]{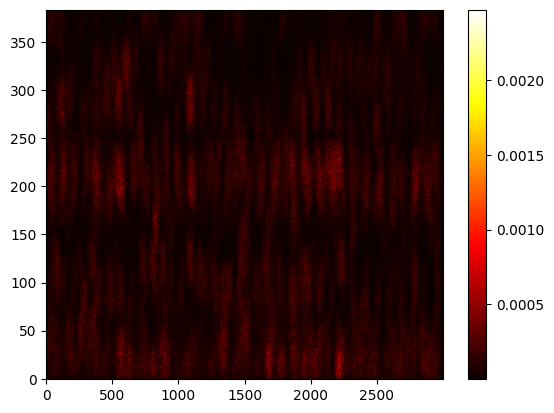}
         \caption{Whisper (frozen)}
         \label{fig:meso_whisper}
     \end{subfigure}
     \begin{subfigure}[b]{0.24\textwidth}
         \centering
         \includegraphics[width=\textwidth]{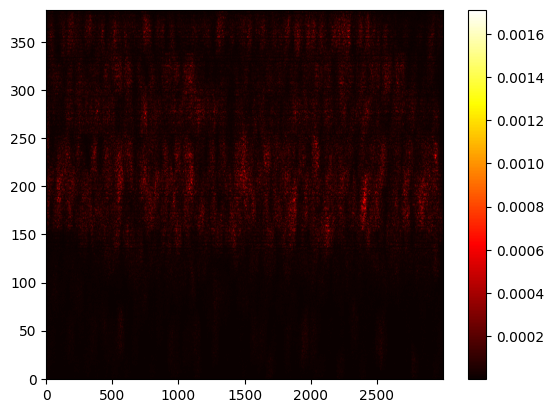}
         \caption{MFCC + Whisper (tuned)}
         \label{fig:meso_tuned}
     \end{subfigure}
        \caption{Saliency maps of front-ends of a bona fide sample for the MesoNet.}
        \label{fig:smap}
\end{figure*}

\begin{figure}[t]
     \begin{subfigure}[b]{0.237\textwidth}
         \centering
         \includegraphics[width=\textwidth]{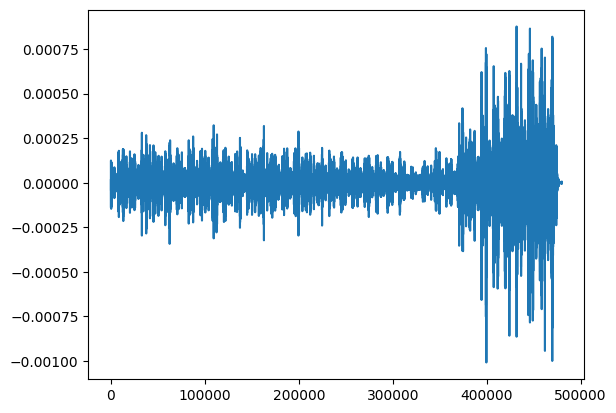}
         \caption{LFCC}
         \label{fig:raw_specrnet_lfcc}
     \end{subfigure}
          \begin{subfigure}[b]{0.22\textwidth}
         \centering
         \includegraphics[width=\textwidth]{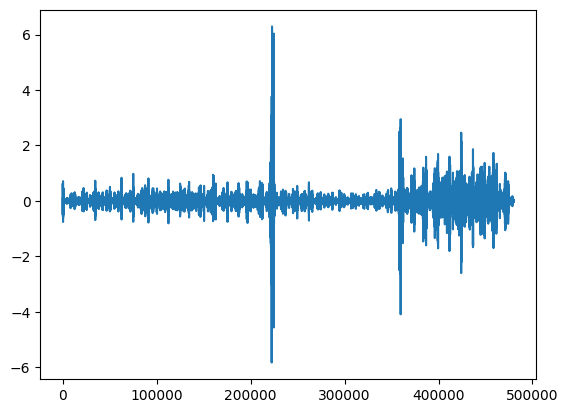}
         \caption{Whisper (frozen)}
         \label{fig:raw_specrnet_whisper}
     \end{subfigure}
        \caption{Gradient calculated on a spoofed signal for the SpecRNet. Note that the y-axis ranges differ.}
        \label{fig:raw_smap}
\end{figure}

\subsection{Whisper fine-tuning}\label{sect:finetuning}

The following benchmark concerned models using Whisper's encoder. This time, however, we did not treat the encoder strictly as a front-end algorithm but rather fine-tuned it to the problem of DF detection. The intuition was the following --- while the features produced by Whisper tend to provide better performance than the typically used front-ends (cf. Tab.~\ref{tab:basic}), this model was trained for a different purpose (ASR), and on the biased (in the sense of DF) data. We assume that the encoder fine-tuned to a specific task, in our case DeepFake detection, might yield even better results. 
For this purpose, we fine-tuned models using the Whisper front-end and evaluated the results with a fine-tuned version of the feature extractor. 
After the initial training presented in Sect.~\ref{sec:baseline_benchmark}, we trained models for additional 5 epochs. This time, however, we unfroze Whisper layers and performed fine-tuning with a learning rate of $10^{-6}$.

One may note that for all architectures (with a notable exception of SpecRNet with Whisper and LFCC features, where we got results worse by less than 9\%), the fine-tuning provided an improvement. Notably, unfrozen Whisper features allowed us to improve even the previous best result --- MesoNet with MFCC features -- by 14.69\%. The best model we obtained, MesoNet with fine-tuned Whisper+MFCC, scored an EER of 0.2672. This surpasses the state-of-the-art results reported in~\cite{wild}, where authors obtained 0.3394 EER evaluating RawNet2~\cite{rawnet2} on DeepFake In-The-Wild after training model on 4s samples from ASVspoof 2019~\cite{asvspoof_2019}. Our results indicate that unfreezing the model and using Whisper extracted features may improve the results of detecting DeepFakes from a significantly different distribution than the set the model was trained on, which would address the vital issue of generalization.

\section{Features comparison}\label{sec:features_comparison}

In order to check if the different front-ends indeed generate different features, we analyze which parts of the input data most significantly affect the detection results. Additionally, we compare two architectures that achieved the highest results  --- SpecRNet containing a recurrent layer (GRU) and MesoNet, which primarily consists of convolutional and max-pooling layers. We use a technique known from adversarial attacks~\cite{madry2018towards} that relies on calculating the gradient on the input data. 

We observed that the choice of the model's architecture has great importance in processing front-ends. As the MesoNet consists of 4 max-pooling layers, the final linear layers of the model receive \textit{max-pooled} information from spatially distributed blocks of size equal to $32\times32$ (see Fig.~\ref{fig:smap}). In turn, information in SpecRNet is processed using the GRU, and the decision is mainly impacted by the end part of the signal (Fig.~\ref{fig:raw_specrnet_lfcc}). Having that said, the models utilizing Whisper features often rely on some characteristics extracted from one or more narrow signal slices (see two peaks around 220k and 360k in Fig.~\ref{fig:raw_specrnet_whisper}). We suppose that Whisper works well with recurrent NN because it extracts \textit{prominent} attributes that do not tend to be hidden, passing through the recurrent sequence.
As we have not assessed any mechanism for spatially-independent processing of the combination of front-ends, our findings suggest a negative impact of Whisper features on other front-ends. Specifically, models that solely rely on MFCC or LFCC for decision-making tend to prioritize the lower band of the front-end and disregard delta and double-delta features (Fig.~\ref{fig:meso_lfcc} and~\ref{fig:meso_mfcc}). Conversely, the 2D Whisper features show the most significant impact of the full band, with a noticeable focus on the middle (Fig.~\ref{fig:meso_whisper}). Fine-tuning Whisper features obtain further performance improvement; surprisingly, in this case, the importance of delta and double-delta features increases (Fig.~\ref{fig:meso_tuned}). We suppose that using the Whisper front-end effectively captures the speech features, and including deltas could enhance the results by providing additional coefficients to describe the spectrum dynamic.

\section{Summary}\label{sec:summary}

In this paper, we show that using Whisper~\cite{Whisper} as a feature extractor in DeepFake detection may improve the efficacy of the detection architectures, particularly in the case of evaluation on samples with significantly different distribution than the training set. We trained the models on ASVspoof 2021 DF and evaluated them on the DeepFakes In-The-Wild dataset.
Using fine-tuned Whisper as a sole feature extractor, we achieved EER of 0.33 $\pm$ 0.01 for all the investigated architectures, which is better than the recent results reported in~\cite{wild}. Moreover, using Whisper and MFCC joint features for MesoNet, we gained EER as low as 0.2672, showing that even \textit{tiny.en} version of Whisper significantly helps in the generalization of DF detection scoring the state-of-the-art results on the In-The-Wild dataset.
In future work, we would like to investigate more Whisper models, following the intuition that larger models used as feature extractors may improve the generalization. Additionally, we are interested in exploring the combinations of front-ends concerning architectures of various models.

\section{Acknowledgements}

This work has been partially funded by Department of Artificial Intelligence, Wroc{\l}aw University of Science and Technology.

\bibliographystyle{IEEEtran}
\bibliography{mybib}

\end{document}